\documentclass[useAMS,usenatbib]{mn2e}

\newcommand{\rev}[1]{{#1}}
\newcommand{\ask}[1]{#1}

\newcommand\epsscale[1]{\def\eps@scaling{#1}}%
\newcommand\plottwo[2]{{%
 \typeout{Plottwo included the files #1 #2}
 \centering
 \leavevmode
 \columnwidth=.95\columnwidth
 \includegraphics[width={\eps@scaling\columnwidth}]{#1}%
 \hfil
 \includegraphics[width={\eps@scaling\columnwidth}]{#2}%
}}%

\usepackage{graphicx,epsf,textcomp,pstricks,psfrag,amsmath,wasysym}

\title[Debris Disc Formation Induced by Planetary Growth]{
Debris disc formation induced by planetary growth 
}

\author[H.~Kobayashi \& T.~L\"ohne]{H.~Kobayashi$^1$\thanks{hkobayas@nagoya-u.jp} and 
T.~L\"ohne$^2$ \\
$^1$ Department of Physics, Nagoya University, Nagoya, Aichi 464-8602,
Japan\\
$^2$ Astrophysical Institute and University Observatory,
Friedrich Schiller University, Schillergaesschen 2-3, 07745 Jena, Germany
} 

\date{Released 2002 Xxxxx XX}

\pagerange{\pageref{firstpage}--\pageref{lastpage}} \pubyear{2013}

\def\LaTeX{L\kern-.36em\raise.3ex\hbox{a}\kern-.15em
    T\kern-.1667em\lower.7ex\hbox{E}\kern-.125emX}

\begin{document}

\label{firstpage}

\maketitle

\begin{abstract}
Several hundred stars older than 10 million years have been observed to
 have infrared excesses. These observations are explained by dust grains
 formed by the collisional fragmentation of hidden planetesimals.  Such
 dusty planetesimal discs are known as debris discs.  In a dynamically
 cold planetesimal disc, collisional coagulation of planetesimals
 produces planetary embryos which then stir the surrounding leftover
 planetesimals. Thus, the collisional fragmentation of planetesimals
 that results from planet formation forms a debris disc.  We aim to
 determine the properties of the underlying planetesimals in debris
 discs by numerically modelling the coagulation and fragmentation of
 planetesimal populations.  The brightness and temporal evolution of
 debris discs depend on the radial distribution of planetesimal discs,
 the location of their inner and outer edges, their total mass, and the
 size of planetesimals in the disc.  We find that a radially narrow
 planetesimal disc is most likely to result in a debris disc that can
 explain the trend of observed infrared excesses of debris discs around
 G-type stars, for which planet formation occurs only before 100 million
 years. Early debris disc formation is induced by planet formation,
 while the later evolution is explained by the collisional decay of
 leftover planetesimals around planets that have already formed.
 Planetesimal discs with underlying planetesimals of radii $\sim
 100\,$km at $\approx 30$\,AU most readily explain the Spitzer Space
 Telescope 24 and 70\,$\micron$ fluxes from debris discs around G-type
 stars.
\end{abstract}


\section{Introduction}

The circumstellar discs observed around several hundred main sequence stars are
mainly gas poor, faint discs, and are mostly revealed by excess infrared
emission around the stars.  Dust grains to which the observed emission
is attributed have lifetimes much shorter than the ages of the
central stars; dust grains are continuously replenished by collisional
cascades from hidden planetesimals, approximately kilometer-sized or larger
bodies.  However, since 
gravity plays a dominant role in determining the outcome of 
planetesimal collisions,
collisional fragmentation of planetesimals to start collisional
cascades needs collisional velocities between planetesimals exceeding their
surface escape velocities. 
Therefore, significant perturbations to increase the collisional
velocities between planetesimals for collisional fragmentation are
required for the formation of debris discs.  Plausible candidates for the
perturbations of planetesimals in debris discs are (i) an early stellar
encounter \citep{kobayashi01}, (ii) the existence of massive gaseous planets
\citep[e.g.,][]{mustill}, and (iii) the formation of planetary embryos in a
planetesimal disc \citep[e.g.,][]{kenyon04}.  In this paper, we focus on
debris discs induced by planetary formation.

Planetesimals formed in a protoplanetary disc are expected to have low random
motions. Collisions between planetesimals 
result in coagulation. Gravitational focusing and dynamical friction
lead to runaway growth of planetesimals and the formation of a single planetary
embryo in each annulus of the disc. Embryos \ask{continue} growing by collisional
accretion of leftover planetesimals, which themselves do not grow
significantly. Once embryos become sufficiently massive, planetesimals start
effective collisional fragmentation. Small bodies resulting from
collisional fragmentation of planetesimals collide with each other and
become smaller still. The collisional cascade grinds bodies down until
radiation pressure from the host star blows them away.

The collisional cascade and the subsequent blow-out reduce the surface
density of planetesimals in the
disc. Along with this reduction, embryo growth stalls
\citep[e.g.,][]{kobayashi10} and the dust mass supplied by the collisional
cascade decreases. On the other hand, the formation timescale of
planetary embryos is longer in the outer disc; planetary growth propagates
from the inner to outer disc. When the dust mass produced by planet
formation decreases in the inner disc, subsequent planet formation increases
the dust production rate in the outer disc. Therefore, 
inside-out planet formation can maintain a detectable amount of debris in
broad planetesimal discs \citep[e.g.,][]{kenyon04,kenyon08}. 
Note that the brightness evolution of debris discs depends on the
broadness of initial planetesimal discs, as shown below.

Planetesimal formation is still a critical issue in the theory of planet
formation. Recently, it has been shown that planetesimals may be formed
from pebbles accumulated in vortices in a turbulent disc
\citep[e.g.,][]{cuzzi} or by direct collisional coagulation of fluffy
dust aggregates \citep{okuzumi}. Planetesimal size depends on the
formation process.  In addition, planetesimal formation might occur in
limited locations.  The initial planetesimal size and spatial
distribution influence the planetary embryo formation timescale and
therefore the temporal evolution of a debris disc.

\begin{figure}
\includegraphics[width=84mm]{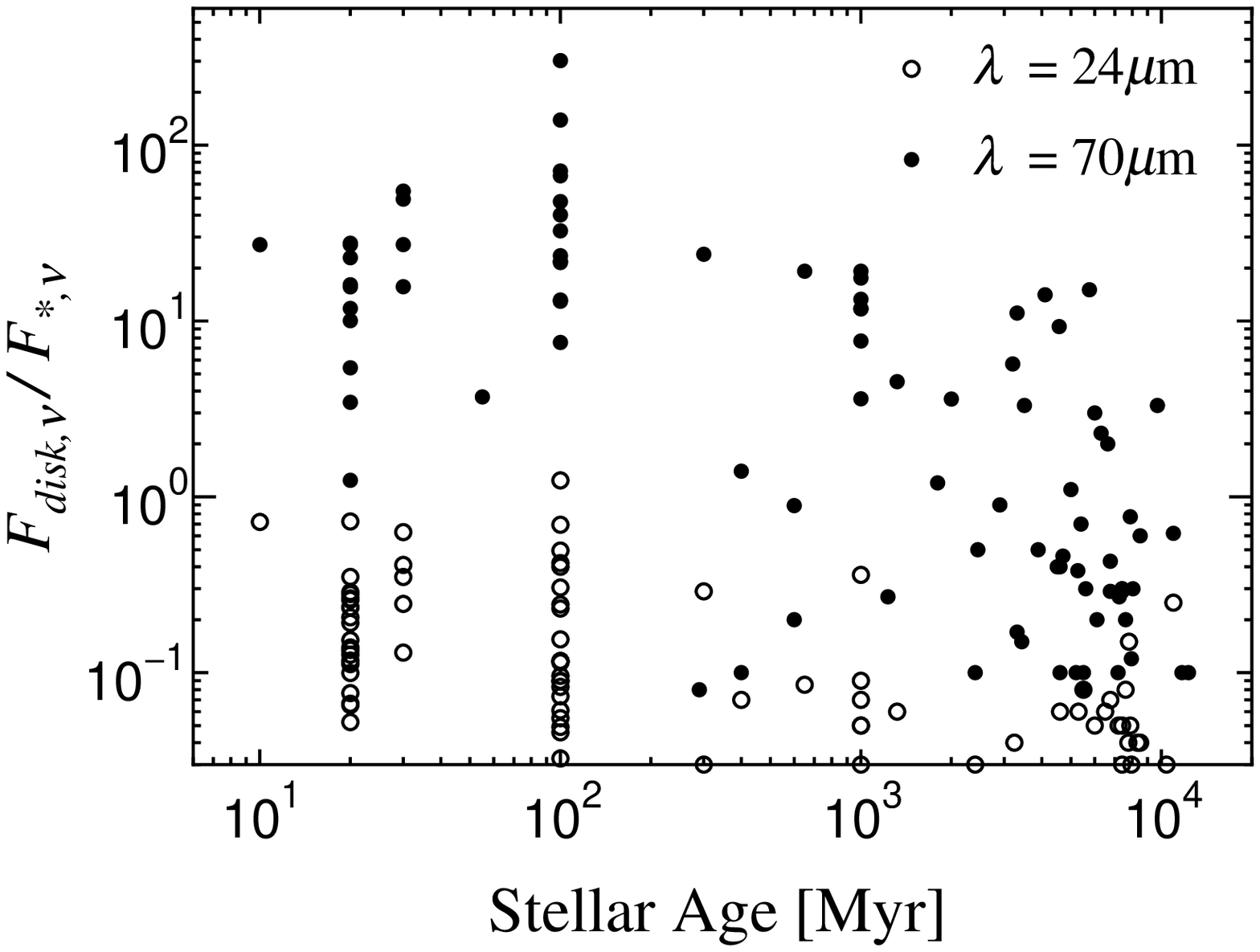} \caption{ The ratios
of disc fluxes $F_{{\rm disc},\nu}$ for G-type stars to the stellar
fluxes $F_{*,\nu}$ observed (but not necessarily detected) at both
wavelengths $\lambda = 24\,\micron$ (open circles) and $70\,\micron$
(filled circles), as a function of the ages of host stars
\citep{bryden,chen05a,chen05b,kim,beichman05,beichman06,meyer,hines,moromartin,hillendrand}.
\label{fig:flux_ratio} }
\end{figure}

Infrared surveys by IRAS, ISO, the Spitzer Space Telescope and others
have shown that infrared excesses from debris discs
around main-sequence stars are common. 
In particular, various photometric surveys 
of hundreds of nearby stars have been conducted by the Spitzer Space
Telescope.  Although observations have been done for many types of stars,
in this paper we are interested in planet formation around solar type
stars, and thus we focus on debris discs around G-type stars.  The observed
infrared excesses mainly decay with stellar age (see
Fig.~\ref{fig:flux_ratio}). 
The time evolution of debris discs induced by planet formation depends
on the radial profiles of the initial planetesimal discs, the initial
sizes of planetesimals, and their total
masses. Therefore, the temporal evolution of infrared excesses
gives constraints on the conditions of planetesimal discs that produce
planets and debris discs. 

In this paper, we investigate the temporal evolution of debris discs
induced by planet formation. In \S~\ref{sc:obs_data}, we estimate the
radii and masses of debris discs from observational Spitzer MIPS data.
In \S~\ref{sc:model}, we describe our collision model and the mass
evolution of bodies through collisions.  In \S~\ref{sc:result}, we carry
out simulations for debris disc formation induced by planetary embryo
formation, and give constraints on initial conditions; the radii of the
inner and outer edges of initial planetesimal disc, initial disc mass,
and the initial radius of planetesimals.  In \S~\ref{sc:discussion}, we
discuss the origin of the inner and outer edges of discs, such as gas
depletion, early stellar encounters, and planetesimal formation, and
their relation with the solar system.

\section{Disc Radii and Masses}
\label{sc:obs_data}

We consider an optically thin debris disc radially distributed from
$r_{\rm in}$ to $r_{\rm out}$. 
The received flux at frequency $\nu$, $F_{{\rm disc},\nu}$, from thermal
emission of the disc around a central star at a distance $D$ from the
observer is given by 
\begin{equation}
 F_{{\rm disc},\nu} = \frac{1}{D^2} \int_{r_{\rm in}}^{r_{\rm out}} 
\int_{\rm s_{\rm min}}^{\rm s_{\rm max}}
2 \pi r 
C_{\rm g} Q_{{\rm abs},\nu} n_{\rm s}(s,r) B_{\nu}(T)  ds dr,\label{eq:flux} 
\end{equation}
where $C_{\rm g} = \pi s^2$ is the geometrical cross section of a
spherical grain with radius $s$, $Q_{{\rm abs},\nu}$ is the absorption
efficiency at frequency $\nu$, $n_{\rm s}(s,r) ds $ is the surface
number density of dust grains with sizes ranging from $s$ to $s+ds$ at a
distance $r$ from the central star, $T$ is the dust temperature
dependent on $s$ and $r$, $B_\nu(T) = 2h\nu^3 / c^2 (e^{h \nu/k_{\rm B}
T} -1)$ is the Planck function, $c$ is the speed of light, $h$ is the
Planck constant, and $k_{\rm B}$ is the Boltzmann constant.  The flux is
sensitive to the smallest grain radius $s_{\rm min}$, compared to the
largest one $s_{\rm max}$. In debris discs, dust grains are supplied
from collisional fragmentation. Small grains with radius $\la
1\,\micron$ are blown out by radiation pressure around the solar type
star \citep[e.g.,][]{burns,kobayashi08,kobayashi09}.

The temperature of dust particles is determined by the energy
equilibrium between stellar radiation and thermal emission, given by
\begin{equation}
 4 \int_{0}^\infty Q_{{\rm abs},\nu} B_\nu (T) d \nu 
  = \frac{R_*^2}{r^2} \int_0^\infty Q_{{\rm abs},\nu} B_\nu (T_*) d \nu,  
\end{equation}
where the stellar radiation is assumed to be blackbody with effective
temperature $T_*$ and $R_*$ is the radius of the central star.  If the
dust radius is much larger than the incident radiation wavelength
$\lambda = c/\nu$, $Q_{{\rm abs},\nu} \approx 1$. For $s \ll \lambda$,
$Q_{{\rm abs},\nu} \approx 2 \pi s k /\lambda$, where $k$ is the
imaginary part of the complex refractive index \citep[e.g.,][]{bohren}.
For blackbody grain ($Q_{{\rm abs},\nu} = 1$), the temperature is
independent of grain size and is given by $T \approx 280 (r/1\,{\rm
AU})^{-1/2} (R_*/R_\odot)^{1/2}\,{\rm K}$, where $R_\odot$ is the solar
radius.

\begin{figure*}
\epsscale{1} \plottwo{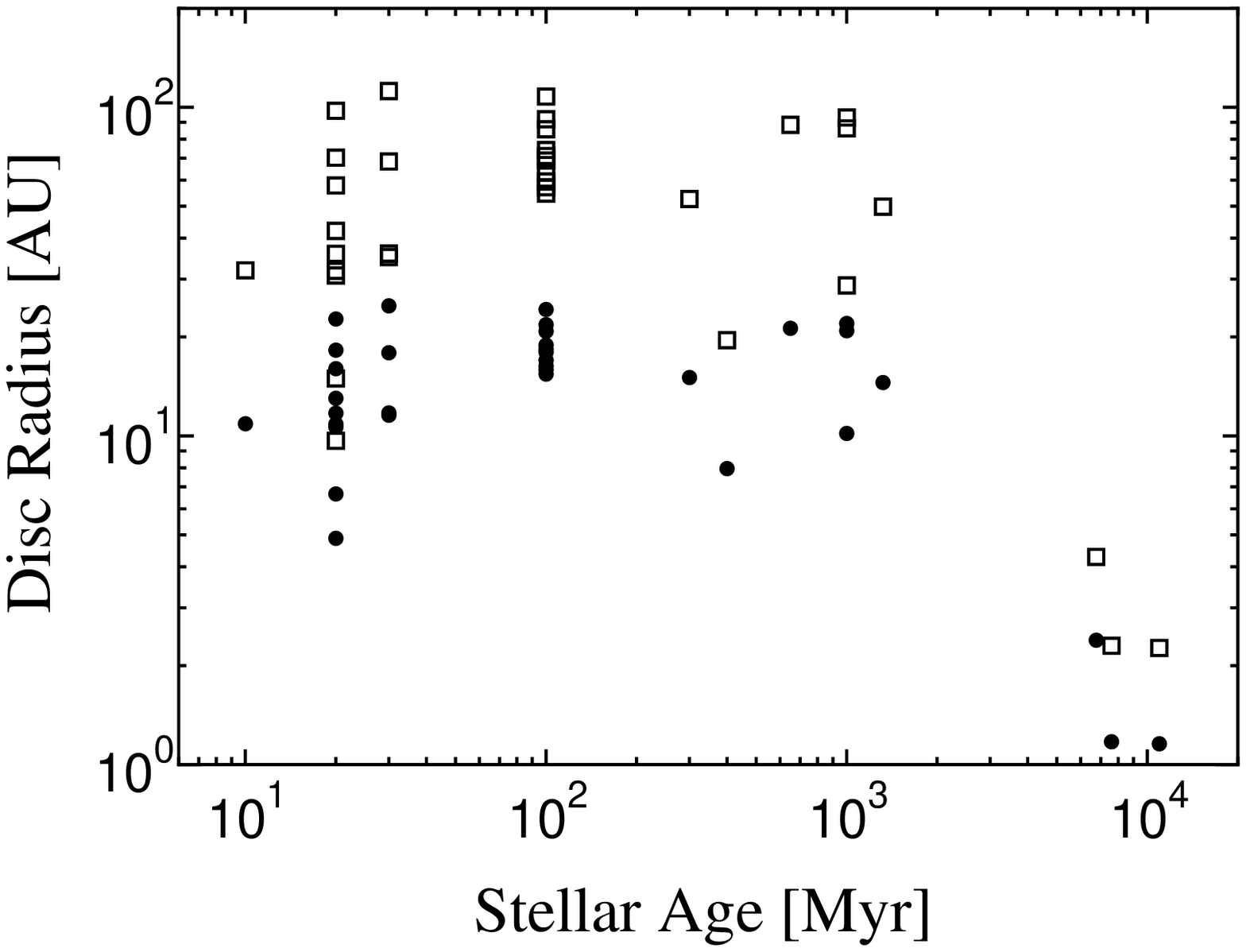}{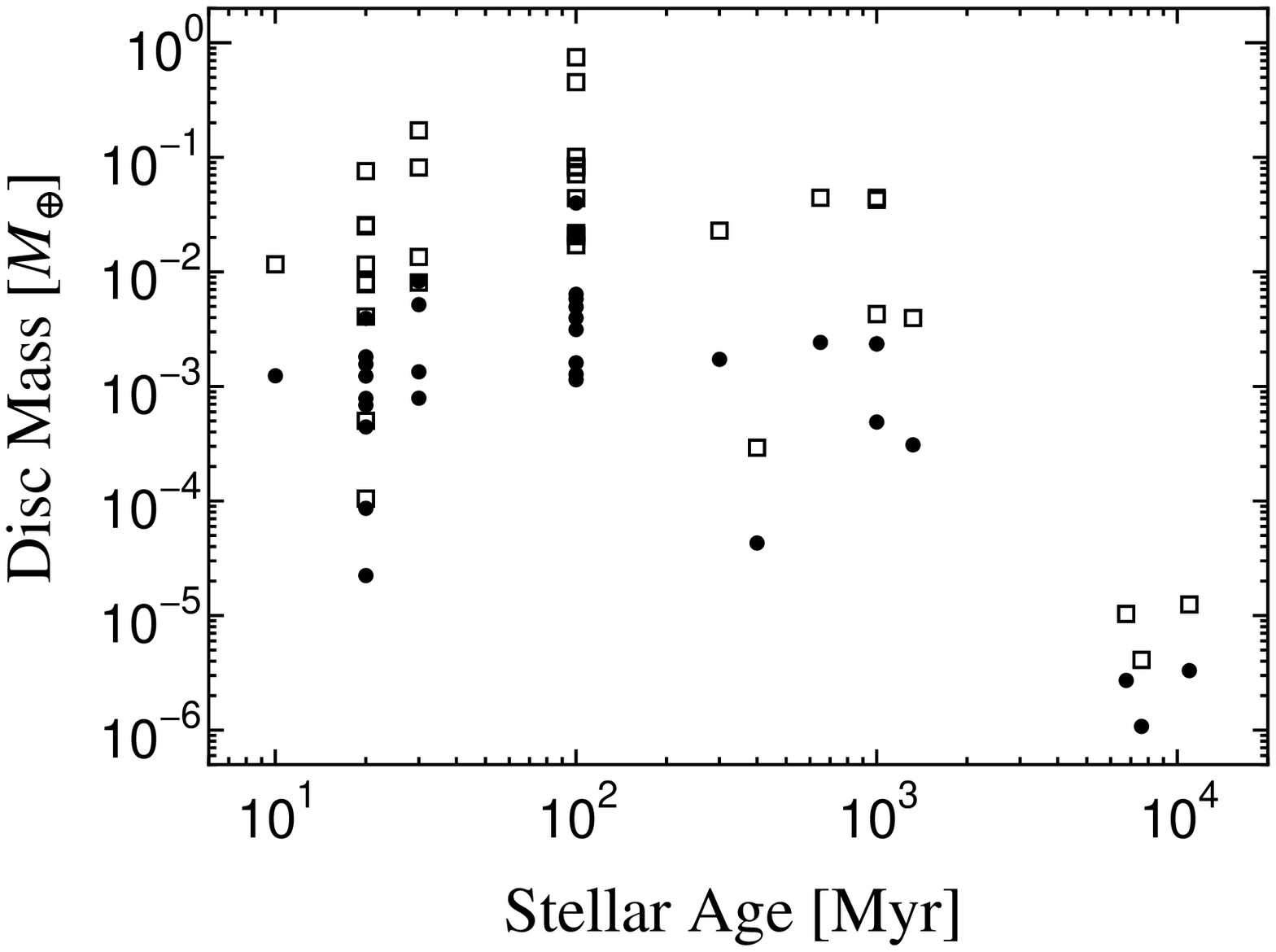}
 \caption{ 
Disc radii and masses obtained from observational data with flux ratio
 larger than 0.06 (0.15) for $\lambda = 24\,\micron$ $(70\,\micron)$ in
 Fig.~\ref{fig:flux_ratio} for dirty-ice grains (open squares) and
 blackbody grains (filled circles), as a function of host star age. 
\label{fig:dist_mass}}
\end{figure*}

The fluxes of debris discs around G-type stars, $F_{{\rm disc},\nu}$,
divided by host star fluxes $F_{*,\nu}$ were obtained from published
observations at 24 and $70\,\micron$ with the MIPS photometer of the
Spitzer Space Telescope (see Fig.~\ref{fig:flux_ratio}).  The debris
discs revealed by high-resolution imaging are narrow rather than broad
and their dust size distributions are approximately given by a single
power law due to collisional cascades.  If we assume that $r_{\rm in} =
r$, $r_{\rm out} = 1.1 r$, $s_{\min}=1\,\micron$, $s_{\rm max} \gg
s_{\rm min}$, and $n_{\rm s}(s,r) = A s^{-7/2}$ where $A$ is a constant,
we obtain $r$ and $A$ from flux ratios at 24 and 70\micron.
Fig.~\ref{fig:dist_mass} shows the radii (r) and masses of observed
disks. The disc masses correspond to the total masses of grains smaller
than 1\,mm.  Although $Q_{{\rm abs},\nu}=1$ for blackbody grains, for
realistic grains, we apply $Q_{{\rm abs},\nu}$ calculated from Mie
theory using the complex refractive index of dirty ice\footnote{ The
composition of dirty ice is ice, organics, and silicates, whose volume
ratio is set to be 2:1:1 according to the local interstellar cloud
\citep{kimura}. The complex refractive index of dirty ice is calculated
using those of ice \citep{warren}, organic refractory material
\citep{li}, and astronomical silicate \citep{draine} through the
Maxwell-Garnett mixing rule.  } \citep{bohren}.  For bodies smaller than
the peak wavelength of the thermal emission spectrum, grain temperatures
calculated for dirty-ice grains are higher than those for blackbody
grains. The disc radii obtained for dirty-ice grains are thus 2--5 times
larger than those for blackbody grains. Disc masses estimated using
dirty-ice grains are larger by an order of magnitude.

The dependence on stellar ages may indicate the evolution of
discs.  Disc radii decrease after several billion years and disc masses peak around 100 million years.  
However, it should be noted that we obtain disc radii and masses
by excluding data with flux ratios smaller than 0.06 (0.15) for 
$\lambda = 24\,\micron$ ($70\,\micron$) due to observational uncertainty
\citep{bryden}. Discs around older stars mainly have flux ratios lower than the
limit for $\lambda = 24\,\micron$ because of large disc radii. 
Therefore, we cannot obtain radii and masses of most old discs through this
analysis.

\section{Model}
\label{sc:model}

 In debris discs, small grains are removed within a short timescale
 mainly by collisional fragmentation and radiation pressure.
 Kilometer-sized or larger parent bodies are required to maintain debris
 discs in timescales comparable to the ages of their host stars.  Parent
 bodies with low random motions initially undergo collisional growth,
 rather than fragmentation, and then planets are formed \ask{via}
 runaway/oligarchic growth. Once planets are substantially massive,
 leftover planetesimals start collisional fragmentation, resulting in
 debris disc formation.  In this section, we describe our model of
 the outcomes of single collisions and the mass evolution of bodies through
 collisions to investigate thermal flux evolution caused by planet
 formation in planetesimal discs.

\subsection{Mass Evolution of Bodies and Collisional Outcomes}

The formation and growth of planetary embryos that induce dust
production in a planetesimal disc are investigated via statistical
simulation developed in \citet{kobayashi10}.  The mass evolution of
bodies is caused by mutual collisions dependent on the random velocities
of bodies, which are affected by the mass distribution of bodies. The
mass and velocity evolution is therefore coupled. Although we do not
follow the orbits of bodies, our simulation that calculates the mass and
velocity evolution can reproduce the embryo formation results of
$N$-body simulation \citep{kobayashi10}.

The random velocities of planetesimals, determined by orbital
eccentricities and inclinations, increase due to stirring by planetary
embryos, which induce collisional fragmentation. Collisions between
already fragmented bodies made them even smaller. Once fragments are
smaller than $\sim 1 \,\micron$ in radius, they are no longer bound to
the host star due to the contribution of radiation pressure and are
mainly blown out before being destroyed in mutual collisions
\citep[e.g.,][]{burns}.  Therefore, such a collisional cascade reduces
the total mass of planetesimals. Although the mass reduction in some
previous studies had been 
investigated under the assumption that catastrophic collisions are
dominant, \citet{kt10} found that weak erosive collisions are more
important for the mass reduction of planetesimals by collisional
cascades. They obtained an analytical solution for the mass reduction
including erosive collisions, and this is reproduced by our
simulations. Therefore, we can accurately follow both planet formation
and debris production.  

Dust production is mainly determined by the total ejecta mass, $m_{\rm
e}$, from a single collision between bodies with masses $m_1$ and $m_2$,
given by 
\begin{equation}
 m_{\rm e} = \frac{\phi}{1+\phi} (m_1+m_2). 
\end{equation}
Here $\phi = m_1 m_2 v_{\rm col}^2 / 2 (m_1 + m_2)^2 Q_{\rm D}^*$ is 
the scaled impact energy, where $Q_{\rm D}^*$ is the specific impact energy required for the
ejection of half of the mass of the colliders and $v_{\rm col}$ is the impact
velocity. 
Collisional simulations of dust aggregates 
(very small particles) suggest that $Q_{\rm D}^*$ is independent of
aggregate size and of order $10^7 {\rm erg/g}$ \citep{wada}. For large
bodies, $Q_{\rm D}^*$ is purely determined by 
gravity and is thus proportional to the square of the surface escape
velocity. For intermediate-sized bodies (1\,cm--100\,km in radius),
hydrodynamic simulations show that $Q_{\rm D}^*$ increases with radius, $s$, for
$s > 0.1$--1\,km,
while $Q_{\rm D}^*$ decreases with $s$ for $s < 0.1$\,km. 
Therefore, we model 
\begin{eqnarray}
 Q_{\rm D}^* &=& Q_{\rm s} \left[ 1 + \left(\frac{s}{1\,{\rm
				     cm}}\right)^{-b_{\rm
 s}}\right]^{-1} 
\nonumber
\\&&
+ 
\rho \left[
Q_{\rm g,1} \left(\frac{s}{1\,{\rm cm}} \right)^{b_{\rm
 g,1}} + Q_{\rm g,2}  \left(\frac{s}{1\,{\rm cm}} \right)^{2} \right], 
\end{eqnarray}
where $\rho$ is material density and $Q_{\rm s}$, $b_{\rm s}$, $Q_{\rm
g,1}$, $b_{\rm g,1}$, and $Q_{\rm g,2}$ are constants. 
We apply $Q_{\rm s} = 1.6\times 10^7 \,{\rm erg/g}$, $b_{\rm s} =
-0.39$, $Q_{\rm g,1} = 1.2 \,{\rm erg\, cm^3/g^2}$, and $b_{\rm g,1} = 
1.26$ based on \citet{benz}, and $Q_{\rm g,2} = 5.0\times 10^{-3}\,{\rm
erg\, cm^3/g^2}$ based on \citet{stewart09}.

We investigate the collisional evolution of bodies in a disc after gas
depletion. Since the eccentricities of planetesimals increase to as much
as unity, we take into account the reduction of the surface density of
planetesimals due to scattering from planetary systems.  According to
the results of $N$-body simulations by \citet{ida}, the eccentricities of
bodies in each mass bin have a Rayleigh distribution. The fraction of bodies with eccentricities larger than unity
is negligible for a small mean value of the distribution, while the
fraction is significant for a mean value close to unity. If the mean
eccentricity becomes larger than 0.25, we remove the fraction of bodies
with eccentricities larger than unity and set a new mean eccentricity
determined by the leftover bodies.

\subsection{Optical Depth and Thermal Emission}

In our simulations, 
we follow the evolution of the optical depth $\tau$, where
\begin{equation}
\tau(r) = \int_{s_{\rm min}}^{s_{\rm max}} C_{\rm g} n_{\rm s} (s,r) ds.  
\end{equation}
Substituting $\tau$ into Eq.~(\ref{eq:flux}), the disc flux is
re-written as 
\begin{equation}
 F_{{\rm disc},\nu} = \frac{1}{D^2} \int_{r_{\rm in}}^{r_{\rm
  out}} 2 \pi r \tau(r) S_\nu(r) dr,
  \label{eq:flux_black}
\end{equation}
where
\begin{eqnarray}
 S_\nu (r) &=& 
\left[
\int_{\rm s_{\rm min}}^{\rm s_{\rm max}}
C_{\rm g} Q_{{\rm abs},\nu} n_{\rm s}(s,r) B_{\nu}(T) ds
\right]
\nonumber 
\\ && \displaystyle \times 
\left[
\int_{s_{\rm min}}^{s_{\rm max}} C_{\rm g} n_{\rm s} (s,r) ds
\right]^{-1}. 
\end{eqnarray}
For blackbody dust, $S_{\nu} (r) = B_{\nu} (T)$ for $T$ corresponding to
the temperature at $r$.  Even for realistic dust, if the size
distribution of $n_{\rm s}(s,r)$ is known, $S_\nu (r)$ can be
obtained. We use $S_{\nu} (r)$ obtained from the temperature of dirty
ice dust and the assumption that $n_{\rm s} \propto s^{-7/2}$.  Although
our simulations yield size distributions, we use the simple $S_\nu(r)$
to save computational cost in fitting over a wide parameter range in \S
\ref{sc:fitting_flux}.  Since there is no significant difference between
the simple and realistic $S_\nu$, our treatment does not affect the
results. \rev{Note that we directly obtain $\tau$ from simulations because the
error of $\tau$ estimated from large bodies under the power-law assumption 
is much greater than that of the simple $S_\nu$. }

Active radial transport of bodies occurs via interaction with gas in
protoplanetary discs or by the Poynting-Robertson effect in gas-free
discs. Hence, simulations with multiple annuli are required. 
However, the radial drift of bodies is negligible in most observed discs 
because the collisional timescale is shorter than the
radial drift timescale due to the Poynting-Robertson effect for $\tau
\ga 3 \times 10^{-5} (r/10 \,{\rm AU})^{-1/2}$ \citep{wyatt}. 
We thus ignore radial transport in our simulations. 

\section{Result}
\label{sc:result}

\subsection{Evolution of Optical Depth}

We simulate the formation of planets and a debris disc at $r =10\,{\rm
AU}$ in a disc with $\Sigma = 0.95\,{\rm g\,cm}^{-2}$, equivalent to the
solid surface density in the minimum-mass solar nebula model
\citep{hayashi}.  Since the size distribution of bodies is narrow prior
to runaway growth of planetesimals, we set a single size
population of planetesimals with radius $s_0 = 10\,$km at the beginning of
the simulation.  In the
simulation, the radius of the smallest bodies is set to be 1\,$\micron$,
below which dust grains are blown out on a Keplerian timescale.

\begin{figure}
\includegraphics[width=84mm]{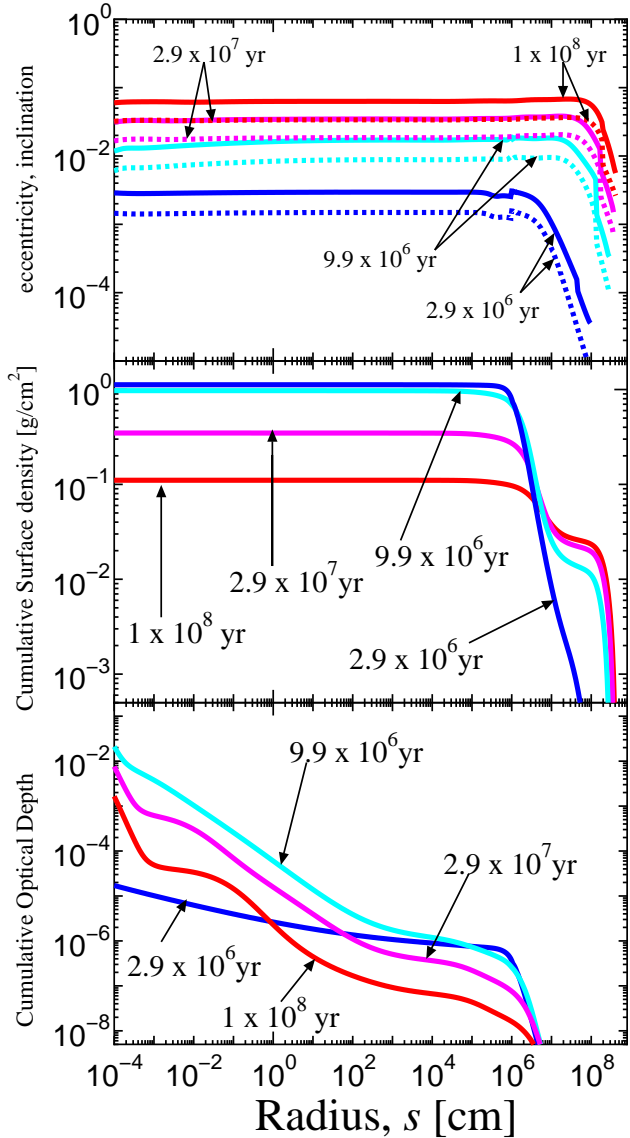} \caption{
 Eccentricity (solid curves, top panel), inclinations (dotted curves,
 top panel), and the cumulative surface density (middle panel) and
 optical depth (bottom panel) of bodies larger than a given radius $s$
 at $r =10\,{\rm AU}$ in a disc of planetesimals with initial radius
 10\,km and the initial surface density $\Sigma = 0.95\,{\rm
 g\,cm}^{-2}$.  \label{fig:mass_dist}}
\end{figure}

The cumulative surface density and optical depth of bodies larger than a
given radius $s$ with eccentricities $e$ and inclinations $i$ are shown
in Fig.~\ref{fig:mass_dist}.  Planetary embryos of radii $\sim 10^3$\,km
form in $\sim 10^7$ years but then the surface density is mainly
determined by bodies of $\sim 10$\,km: Runaway growth produces massive
embryos but most remaining bodies retain the initial size almost without
growth.  The stirring of massive embryos increases $e$ and $i$ of
leftover planetesimals and induces their collisional fragmentation. The
collisional cascade of bodies smaller than 10\,km reduces the surface
density and optical depth of the disc within a timescale of several
$10^7$\,years.

\begin{figure}
\includegraphics[width=84mm]{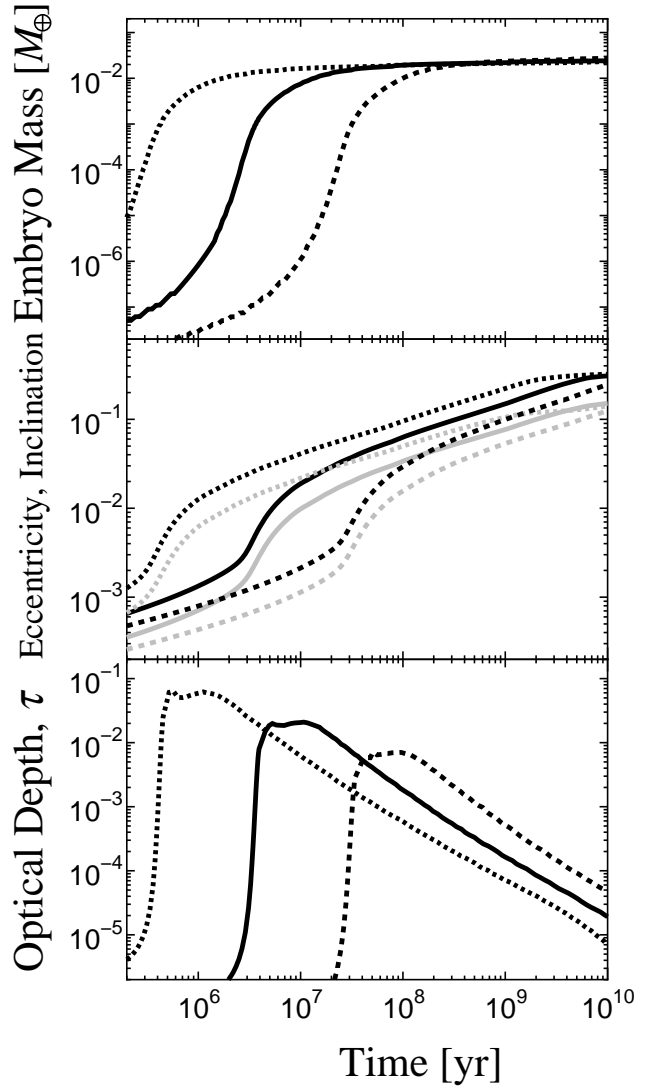} \caption{
 Evolution of planetary embryo mass (top panel), eccentricity (black)
 and inclination (gray) of 10\,km-radius bodies (middle panel), and
 optical depth $\tau$ (bottom panel), starting from 10\,km-radius
 planetesimals for $\Sigma = 2.7\,{\rm g\, cm}^{-2}$, $r = 5\,{\rm AU}$
 (dotted curves), $\Sigma = 0.95\,{\rm g\,cm}^{-2}$, $r =10\,{\rm AU}$
 (solid curves), and $\Sigma = 0.34\,{\rm g\,cm}^{-2}$, $r = 20\,{\rm
 AU}$ (dashed curves), where $\Sigma$ is the solid surface density.
 \label{fig:tau_evo}}
\end{figure}

Figure \ref{fig:tau_evo} shows the evolution of planetary embryo
mass\footnote{We define planetary embryo mass as the average mass of
``runaway bodies'' that can have orbital separations of 10 Hill radii \citep[see][]{kobayashi10}.}
and optical depth $\tau$ at 5, 10, and 20\,AU. Planetary embryos
initially grow exponentially (runaway growth) but subsequently they grow
slowly due to the high $e$ and $i$ of stirred planetesimals (oligarchic
growth). Once embryo masses reach $\sim 10^{-3} M_\oplus$, the optical
depth rapidly increases due to collisional fragmentation of
planetesimals induced by embryo formation.  Finally, their growth is
stalled due to the reduction of the surface density of planetesimals
caused by collisional cascade, which gradually reduces the optical
depth.

Debris disc formation induced by planet formation was also investigated
in previous studies \citep{kenyon04,kenyon08,weidenschilling}.  Due to
high computational costs, the radii of the smallest bodies in their
simulations were set to be much larger than the blow-out size. The 
authors mainly determined the evolution of the resultant smaller bodies based
on the theory of collisional cascades (power law distribution). However,
the distribution does not follow the power law distribution before the
onset of active collisional fragmentation due to planet formation and
even for collisional equilibrium the distribution has wavy structures around the
blow-out size \citep{lohne}. To calculate optical depth accurately,
bodies larger than the blow-out size should be followed in
a simulation. Therefore, we treat all bodies larger than the blow-out size
(1\,$\micron$ in radius).

\begin{figure}
\includegraphics[width=84mm]{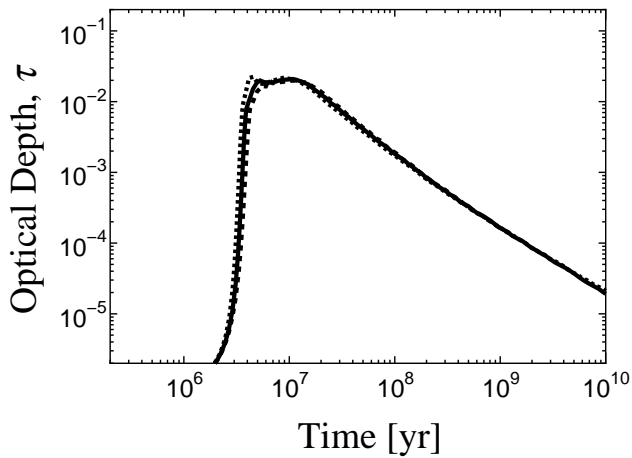} \caption{ Same as
 the bottom panel of Fig.~\ref{fig:tau_evo}, but optical depth $\tau$
 and time $t$ are divided by $(\Sigma / 0.95 \, {\rm g\, cm}^{-2})$ and
 $(\Sigma / 0.95 \, {\rm g\, cm}^{-2})^{-1} (r/10\,{\rm AU})^{3/2}$,
 respectively.  \label{fig:scaled}}
\end{figure}

The rapid increase in $\tau$ is caused by embryo formation via runaway
growth.  We empirically know that the growth timescale is inversely
proportional to $\Sigma \Omega$ in runaway growth
\citep[e.g.,][]{ormel}, where $\Omega$ is the Keplerian frequency. On
the other hand, $\tau$ gradually decreases on a collisional timescale,
that is inversely proportional to $\Sigma \Omega$ \citep{kt10}. The
optical depth is expected to be proportional to $\Sigma$. Therefore, we
scale the temporal evolution of $\tau$. As shown in
Fig.~\ref{fig:scaled}, the scaled temporal evolution is in good
agreement with the actual simulation. Owing to this scaling, numerous
simulations with different annular radii and surface densities are not
necessary to treat a broad disc: We can calculate evolutionary fluxes
for each choice of disc parameter based on the time evolution of $\tau$
given by a single simulation.

\begin{figure}
\includegraphics[width=84mm]{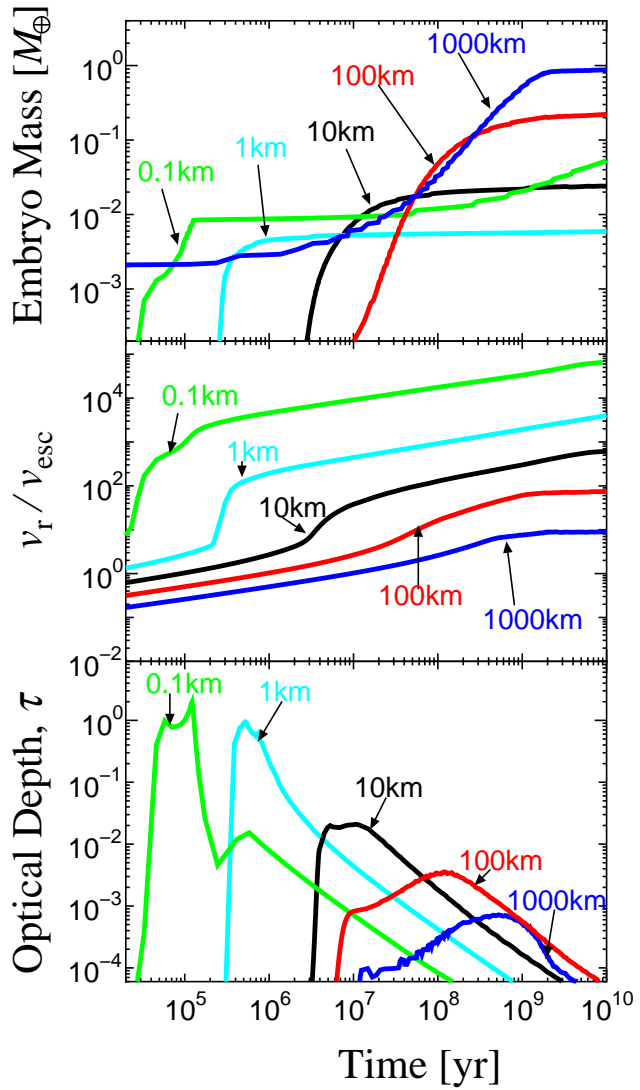}
 \caption{ 
Temporal evolution of embryo mass (top), the ratio of random
 velocity $v_{\rm r}$ of bodies with initial radii $s_0$ to their surface escape
 velocity $v_{\rm esc}$ (middle), and optical depth (bottom) for
 different initial planetesimal radii $s_0 
 = 0.1$--$1000$\,km with $\Sigma = 0.95\, {\rm g\, cm}^{-2}$ at 10\,AU. 
\label{fig:tau_size_dep}}
\end{figure}

It should be noted that the temporal evolution of $\tau$ depends on
the initial planetesimal radius $s_0$, as shown in
Fig.~\ref{fig:tau_size_dep}.  For smaller $s_0$, runaway growth occurs
earlier and thus $\tau$ increases earlier.  Once active dust production
occurs, \ask{more frequent collisions of more fragile planetesimals (lower
$Q_{\rm D}^*$) for smaller $s_0$ 
lead to higher dust production: Smaller $s_0$ results in higher $\tau$,
whereas $\tau$ decays earlier for smaller $s_0$.}  For $s_0 \la
10\,$km, $\tau$ increases along with runaway growth of
planetesimals, while $\tau$ increases prior to runaway growth for
larger planetesimals.  When the runaway growth starts, planetesimals
have random velocities $v_{\rm r} = \sqrt{e^2 + i^2} v_{\rm k}$ as large
as the surface escape velocity, $v_{\rm esc}$, of the planetesimals,
where $v_{\rm k}$ is the Keplerian velocity. Since their specific impact
energies at the beginning of runaway growth are much smaller than
$Q_{\rm D}^*$ of planetesimals with $s_0 \la 10\,$km, $\tau$ is very low
before runaway growth and suddenly increases at the onset of runaway
growth. For $s_0 \ga 100$\,km, the specific impact energy is slightly
smaller than or comparable to $Q_{\rm D}^*$ of initial planetesimals at
the beginning of runaway growth.  Due to collisional erosion of
planetesimals, collisional cascades increase $\tau$ prior to runaway
growth, resulting in the small peak value of $\tau$.  Therefore, the
dependence of $\tau$ evolution on $s_0$ is complicated, because
fragmentation efficiency of planetesimals determined by $Q_{\rm D}^*$
depends on $s_0$. Although we cannot derive a formula for $s_0$
dependence, the scaling for $\Sigma$ and $r$ is valid for each value of
$s_0$.

\subsection{Evolution of Disc Flux
\label{sc:fitting_flux}
}

We obtain the disc flux by the integration of $\tau$ over $r$ using 
Eq.~(\ref{eq:flux_black}). For the integral, we introduce a power-law
surface density model of 
\begin{equation}
 \Sigma = x \Sigma_{10} \left(\frac{r}{10\,{\rm AU}}\right)^{-p},
\end{equation}
where $\Sigma_{10} = 0.95\, {\rm g\, cm}^{-2}$ is the reference surface
density at 10\,AU, corresponding to that at 10\,AU in the minimum mass
solar nebula model \citep{hayashi}, and $x$ is a scaling
parameter. 

\begin{figure}
\includegraphics[width=84mm]{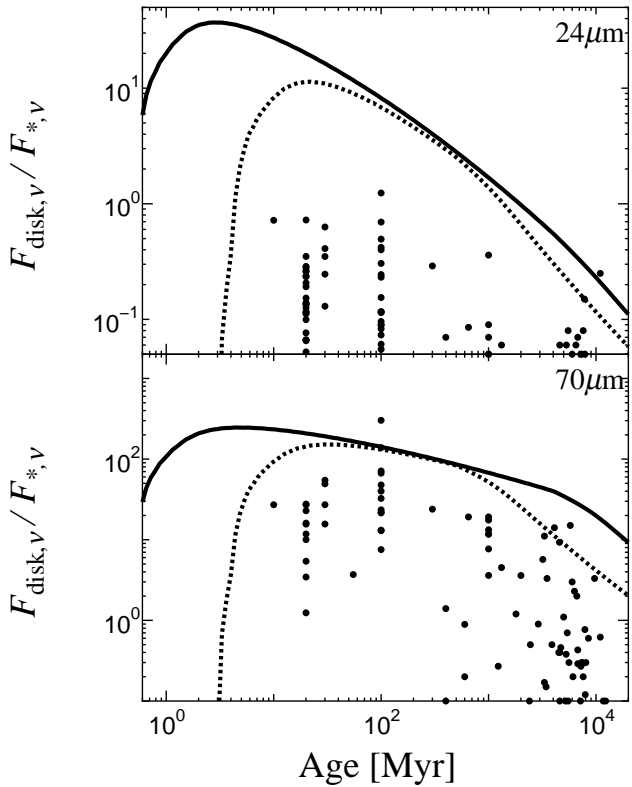} \caption{ Temporal evolution
 of flux ratios at $24\,\micron$ (top) and $70\,\micron$ (bottom) for
 $s_0 = 10$\,km, $x = 1$, and $p=1.5$ in a wide disc ($r_{\rm in} =
 5$\,AU and $r_{\rm out} = 100\,$AU; solid curves) and a narrow disc
 ($r_{\rm in} = 10$\,AU and $r_{\rm out} = 50\,$AU; dotted
 curves). Circles indicate the observational data shown in
 Fig.~\ref{fig:flux_ratio}.  \label{fig:evo_bright}}
\end{figure}

Figure~\ref{fig:evo_bright} shows the temporal evolution of flux from a
wide disc from 5 to 100\,AU and from a narrow disc from 10 to 50\,AU
with $x=1$ and $p = 1.5$, initially composed of 10\,km-radius
planetesimals ($s_0 = 10$\,km).  The fluxes increase at around $10^6$
years for the wide disc, but 10 times later for the narrow disc.
This is caused by planet formation around the inner edges of the
discs. The vertical optical depth around the inner edges gradually
decreases after planet formation occurs (see Fig.~\ref{fig:tau_evo}),
while planet formation propagates to the outer disc. This growth
propagation maintains high fluxes until the growth front reaches the
outer, cold disc or the outer edge of the disc. For the $24\,\micron$
flux, thermal emission from the disc beyond several 10\,AU has a smaller
contribution. Since the growth front reaches the cold region, the flux
decreases after 100 million years. On the other hand, $70\,\micron$
thermal emission from \ask{cold disc} has a larger contribution. The
flux decreases when planetary formation has finished at the outer edge
of disc.

\begin{figure}
\includegraphics[width=84mm]{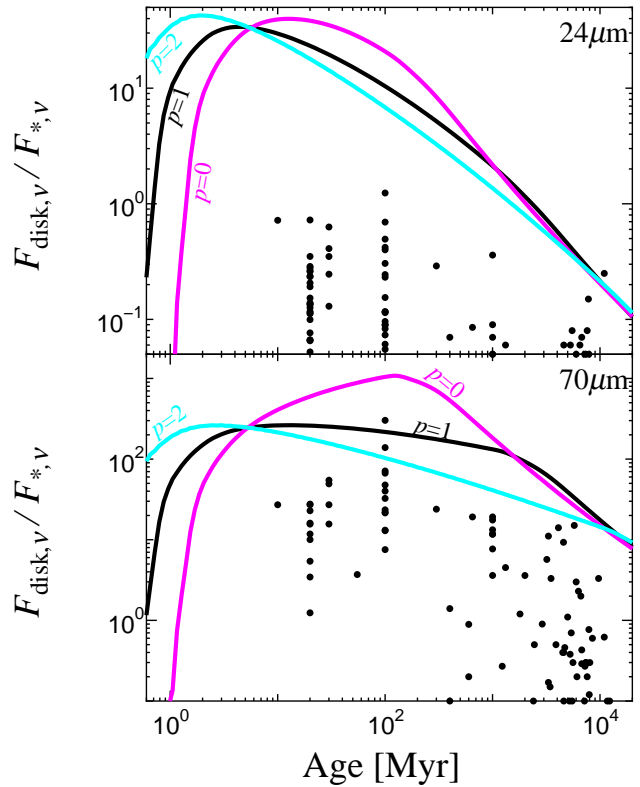}
 \caption{ 
Dependence of flux evolution on $p$ for  $s_0 = 10$\,km, $x = 1$, 
$r_{\rm in} = 5$\,AU and $r_{\rm out} = 100\,$AU. 
\label{fig:evo_bright_dep_p}}
\end{figure}

Fig.~\ref{fig:evo_bright_dep_p} shows the flux ratio evolution for
different radial distributions of surface density. Since the surface
densities at the inner edge are different, the flux
ratios rise later for large $p$. After the rapid rise of flux
ratios, growth propagation in the disc determines the flux ratio
evolution. For $70\,\micron$, the flux ratios decrease for $p\geq 1$, while
the ratio increases for $p=0$. 

\begin{figure}
\includegraphics[width=84mm]{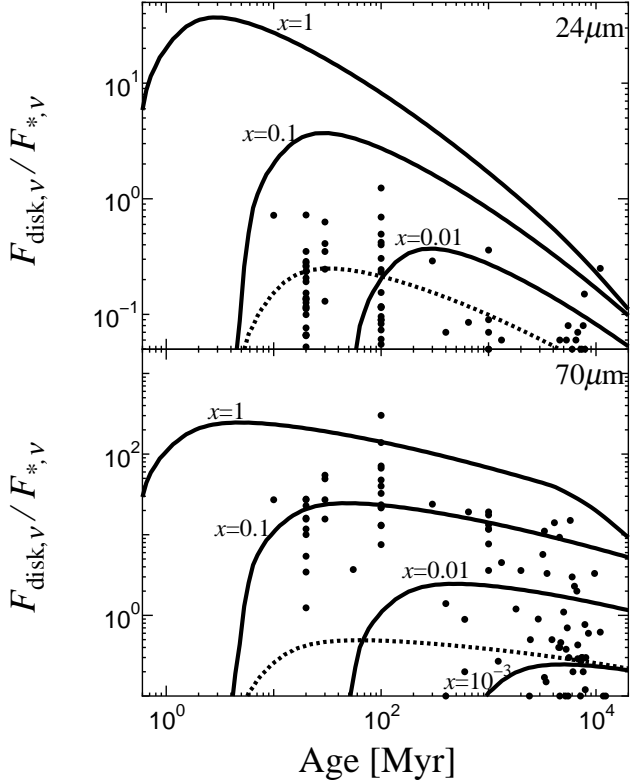}
 \caption{ 
Dependence of flux evolution on $x$ ranging from $10^{-3}$ to 1 for
 $s_0 = 10$\,km, $p = 1.5$,  
$r_{\rm in} = 5$\,AU, and $r_{\rm out} = 100\,$AU (solid curves). 
For $x=10^{-3}$, the flux ratio at $\lambda = 24\,\micron$ is below 0.05. 
Since the planet formation timescale becomes longer for low $x$, we also apply
 a smaller inner disc radius of $r_{\rm in} = 1$\,AU for $x=1\times 10^{-3}$
 (dotted curve). 
\label{fig:evo_bright_dep_x}}
\end{figure}

Since the flux ratios from discs with $x=1$ are much larger than those
obtained from observations, we further investigate the temporal
evolution of flux ratios for less massive discs
(Fig.~\ref{fig:evo_bright_dep_x}).  The maximum flux ratios decrease
with decreasing $x$, while flux ratios rise later for smaller $x$
because of a longer planet growth timescale.  In order to explain lower
observational flux ratios at 10--100 million years, small $x$ and
$r_{\rm in}$ are necessary: The flux evolution for $x=10^{-3}$ and
$r_{\rm in} = 1$\,AU seems more reasonable. However, the model fluxes
are relatively high at $24\,\micron$ and too low at $70\,\micron$,
compared to observational data: The model disc radius is smaller than
those estimated from the observational data. Indeed, while the disc
radii estimated from observational data are much larger than 5\,AU at
10--100 million years (see Fig.~\ref{fig:dist_mass}), the model flux
with $x=10^{-3}$ comes from the disc inside 5\,AU before 1 billion
years.

\begin{figure}
\includegraphics[width=84mm]{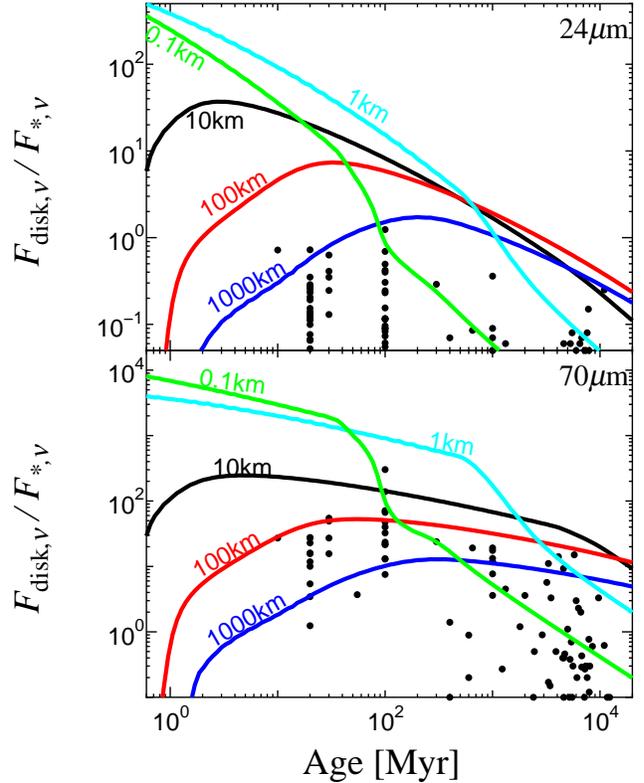} \caption{ Dependence of
 flux evolution on $s_0$ ranging from $0.1$ to 1000\,km for $x=1$, $p =
 1.5$, $r_{\rm in} = 5$\,AU, and $r_{\rm out} = 100\,$AU.
 \label{fig:evo_bright_dep_r0}}
\end{figure}

The temporal flux evolution depends on the initial size of planetesimals
(Fig.~\ref{fig:evo_bright_dep_r0}).  For small planetesimals, planetary
formation starts at the inner edge early and, of course, the growth
front reaches the outer edge early.  For $s_0 = 0.1$ and $1$\,km, the
fluxes are very high before 10 million years and then start decreasing at
$10^8$--$10^9$ years: These model fluxes are much larger than
observational ones before $10^9$ years. 
On the other hand, the fluxes increase later for
larger planetesimals and the fluxes remain high even after $10^9$
years: The model fluxes significantly exceeds observational ones after
$10^8$ years. 
Therefore, the smooth discs seem unlikely to explain the
observational data even for a wide range of initial planetesimal sizes
(0.1--1000\,km).

\subsection{A Typical Disc}

For each initial planetesimal radius, the most likely parameters to
explain the observational data are found using a Monte-Carlo method for
$x=10^{-4}$--100, disc edge radii $r_{\rm in}$ and $r_{\rm out}$ ranging
from 5\,AU to 400\,AU, and $p = 0$--2. From $\chi^2$ tests, we obtain
the best-fit parameters for the averaged data of the flux ratios at 24
and 70\,$\micron$ and their ratios in age bins\footnote{The mean
values are $(\log_{10} {\rm Age [yr]}, \log_{10} F_{\rm
disc,\nu}/F_{*,\nu}|_{24\micron},$ $\log_{10} F_{\rm
disc,\nu}/F_{*,\nu}|_{70\micron}) = (7.32,-0.63,1.17), (8.03, -0.73,
1.55)$, $(9.02, -1.27,0.70)$, and $(9.80,-1.39,-0.51)$, which are
obtained from the average of logarithmic values for age bins, $\log_{10}
{\rm Age} [{\rm yr}] = 6.5$--7.5, 7.5--8.5, 8.5--9.5, and
9.5--10.5. For averaging, we use all data if the flux ratios
exceed 0.15 at $70\,\micron$, and we adopt 0.03 for the flux ratios at
$24\,\micron$ if they are smaller than 0.06.}.  Note that our best fit
disc is no attempt to cover the observed variety of disc radii and
masses with a single set of disc parameters. Instead it is meant to
represent a typical or average disc.

\begin{figure}
\includegraphics[width=84mm]{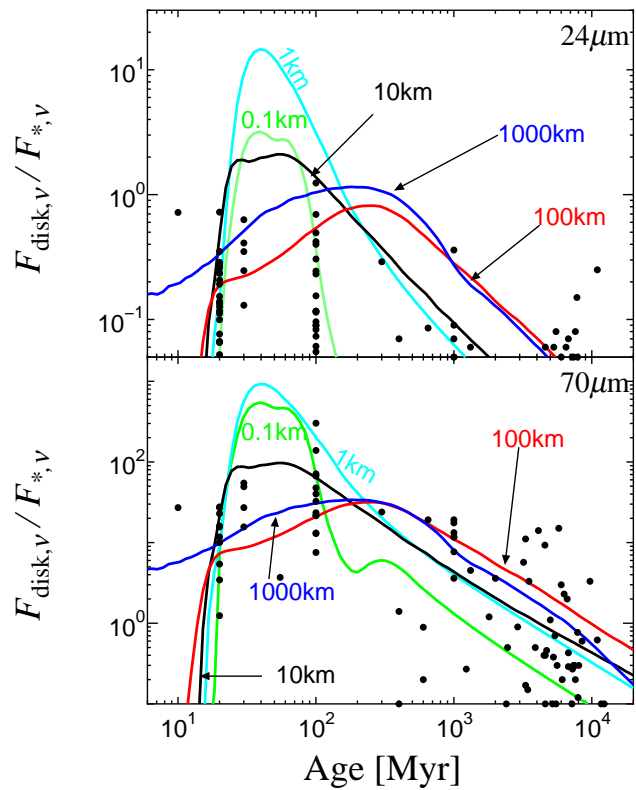}
 \caption{ 
 Temporal evolution of the flux ratios using best-fit parameter sets listed in Table~\ref{tab:best_fit} for initial planetesimal radii of 0.1, 1,
 10, 100, 1000\,km. 
\label{fig:best_fit}}
\end{figure}

The likely parameter sets for several values of the initial planetesimal
radius $s_0$ are listed in Table~\ref{tab:best_fit}.
Fig.~\ref{fig:best_fit} shows the flux-ratio evolution for these likely
parameters.  Narrow discs are most likely to reproduce the observed
fluxes.  Due to the narrowness, the $\chi^2$ values are similar for
broad ranges of $p$ and $x$ as long as the total planetesimal mass
$M_{\rm tot}$ is the same: The best fits are achieved for the relation
$M_{\rm tot} = 2\pi x \Sigma_{10} (10{\rm \,AU})^p (r_{\rm out}^{2-p}
-r_{\rm in}^{2-p})/(2-p)$. The errors for $p$ and $x$ are not described
in Table~\ref{tab:best_fit} because of correlation. They can be
evaluated using the above relation.  The smallest $\chi^2$ values are
obtained for $s_0 = 0.1{\rm \,km}$, which is achieved in a very narrow
parameter space. This is caused by the absence of fitted data at around
30--90 million years. If we exclude this parameter space, cases with
$s_{0} = 100\,{\rm km}$ best represent observations. For $s_{0} =
100\,$km the $\chi^2$ values near minimum are achieved for the wide
range of $M_{\rm tot}$ (see Table \ref{tab:best_fit}). Therefore, the
most likely initial planetesimal size seems to be of order 100\,km.

\begin{table*}
\caption{ The best-fit parameters to observational data: For each
initial planetesimal radius $s_0$, the inner and outer radii of the disc
$r_{\rm in}$ and $r_{\rm out}$, the surface density scaling factor $x$,
the surface density radial slope $-p$, and the total mass $M_{\rm tot}$
of the initial planetesimal disc are listed. The errors for $x$ and
$p$ are discussed in the main text. }  \label{tab:best_fit} \centering
 \begin{tabular}{c | c c c c | c}
 \hline
$s_0$ [km]  & $r_{\rm in}$ [AU]& $r_{\rm out}-r_{\rm in}$ [AU] & $x$ &
  $p$ & $M_{\rm tot}$ [$M_\oplus$]\\
  \hline
 0.1  & $70.6_{-0.04}^{+1.3}$   & $13.5_{-5.1}^{+0.03}$ & 0.23 &  1.1 & $5.7_{-0.2}^{+0.5}$\\
  1 & $39.6_{-0.8}^{+0.7}$ & $1.47_{-0.77}^{+1.33}$ & 0.40 & 0.56 & $14_{-1.5}^{+0.01}$\\
10 & $34.3_{-0.8}^{+0.1}$ & $2.23_{-0.12}^{+1.53}$ & 1.34 & 0.04 & $33_{-1.8}^{+2.9}$\\
100 & $30.5_{-1.3}^{+0.7}$ & $1.75_{-0.07}^{+1.48}$ & 2.6 & 0.0 & $45_{-13}^{+49}$\\
1000 & $26.3_{-1.2}^{+0.4}$ & $1.16_{-0.43}^{+1.29}$ & $79$ & $1.9$ & $122_{-71}^{+50}$\\
    \hline 
\end{tabular}
\end{table*}

\section{Discussion}
\label{sc:discussion}

Planetary embryos formed via runaway growth in a planetesimal disc
induce collisional fragmentation of planetesimals that can supply dust
grains to a debris disc. Planet formation propagates \ask{from the inner
to outer disc}, and the dust supply region moves in turn with planet
formation. Since collisional fragmentation reduces the surface density
of planetesimals, the dust production rate gradually decreases after
planet formation occurs.  As a result, planet formation forms narrow
debris discs from inside out if initial planetesimal discs have a smooth
radial distribution or have no radial cutoff.  However, we find that
planetesimal discs are likely narrow to account for most observational
fluxes of debris discs around G-type stars. Planet formation in narrow
planetesimal belts can also better explain debris discs around A-type
stars \citep{kennedy}.  For G stars, the plausible disc radii are
estimated in this paper to be 25--35\,AU for an initial planetesimal
radius of $s_0 \ga 10$\,km, whereas they are 40--80\,AU for smaller
initial planetesimals. Their radii are comparable to those obtained from
the simple estimate shown in \S~\ref{sc:obs_data}, whereas the total
mass of bodies is not well determined from the simple estimate because
of the uncertainty of the size distribution of large bodies (see
Fig.~\ref{fig:dist_mass} and Table \ref{tab:best_fit}).  In addition,
from more detailed analyses we find that 100\,km sized planetesimals are
most likely to explain the observational fluxes.

The fluxes of debris discs are mainly determined by thermal emission of
dust grains of size 1--100\,\micron.  However, in previous studies, the
debris disc fluxes caused by planet formation are calculated by
extrapolation from the results of simulations that only treat bodies
larger than $\sim 1\,$m \citep{kenyon04,kenyon08}.  Therefore, their
model fluxes are underestimated by a factor 3--10, compared to
simulations handling down to $1\,\micron$-sized dust grains.  On the
other hand, previous simulations included a gaseous component with a
lifetime of $10^7$ years, while we ignore the effects of gas even in
early times. Even starting with similar initial planetesimals, the time
when disc fluxes increase is earlier for our simulation. This means that
gas drag plays an important role in flux evolution before 100 million
years.  Indeed, since collisional cascades cannot effectively produce
$1\,\micron$-sized bodies in a gaseous disc \citep{kobayashi10}, disc
fluxes may not rise before gas depletion occurs.

If gas depletion is taken into account, the inner edge of a disc may
also be naturally explained.  If discs still have as much gas as the
minimum-mass solar nebula, collisional fragmentation between bodies
smaller than about 10\,m does not occur because of damping due to gas
drag. Hence, collisional cascade induced by planet formation do not
significantly produce bodies smaller than 10\,m and the bodies at the
low-mass end of the collisional cascade are depleted by radial drift due
to gas drag \citep{kobayashi10,kobayashi11,kobayashi12}.  After gas
depletion, collisional cascades can produce smaller bodies and then
debris disc fluxes increase.  If planet formation occurs inside a radius
$r_{\rm gas}$ prior to gas depletion, planets are subsequently formed
beyond $r_{\rm gas}$ in a gas-free disc. If $r_{\rm gas}$ corresponds to
the inner edge radius $r_{\rm in}$ of a planetesimal disc that we assume
in this paper, the flux evolution is expected to be similar: The inner
edges of planetesimal discs may be related to the gas depletion of
protoplanetary discs. On the other hand, gas giant planets formed inside
$r_{\rm in}$ dynamically clean up around their orbits, which may also
contribute to the formation of the inner edges of planetesimal discs.

Planetesimal formation beyond several 10\,AU is difficult because
radial drift is more rapid than the collisional growth of dust in
protoplanetary discs; possible formation region of planetesimals via
collisional growth is inside several 10\,AU \citep{okuzumi}. 
On the other hand, if stars are born in a cluster, stars experience
close stellar encounters during their escape from the cluster. Such an
early stellar encounter truncates the disc; a stellar passage at
$\approx 100\,$AU as expected in the relatively dense clusters produces
an outer edge of planet forming region at 30\,AU \citep{kobayashi01} and
explains some dynamical properties of Kuiper belt objects
\citep{kobayashi05}.  Therefore, planetesimal formation and/or an early
stellar encounter may explain the outer edges of planetesimal discs.

In the solar system, Jupiter and Saturn were formed before the gas
lifetime of the solar nebula, while Uranus and Neptune were formed after
significant gas depletion. Planet formation in gas-free discs might have
occurred beyond 10--20\,AU. An early stellar encounter as an explanation
of the orbital distribution of Kuiper belt objects may yield the outer
edge of the planetesimal disc at around 50\,AU
\citep{kobayashi05}. Therefore, narrow planetesimal discs composed of
large planetesimals ($s_0 \sim 100$\,km) that are very likely to form
debris discs might be similar to that of the solar system.  Taking into
account gas depletion, the similarity to the solar system should be
address in future studies.

\vspace{0.5cm}

We thank A. Mustill for helpful comments that helped to improve 
our manuscript. HK gratefully acknowledges support from 
Grants-in-Aid from MEXT (23103005). TL acknowledges support from the Deutsche Forschungsgemeinschaft, grant Lo 1715/1-1.

\end{document}